\documentclass[prd,showkeys,floatfix,twocolumn,amsmath,amssymb,floatfix]{revtex4}
\usepackage{graphicx,color,dcolumn,booktabs,bm}
\usepackage{longtable,lscape}
\usepackage{amssymb}
\usepackage{indentfirst}
\usepackage{epsfig}
\usepackage{feynmf}
\usepackage{epstopdf}
\usepackage{slashed}
\usepackage{cases}
\usepackage{multirow}
\usepackage[colorlinks,citecolor=blue,anchorcolor=red,menucolor=red,linkcolor=red,filecolor=red,runcolor=red,urlcolor=blue,frenchlinks=red]{hyperref}

\newcommand{\feynp}[1]{#1\kern-0.45em/}
\def\dab{\int^{\alpha_{max}}_{\alpha_{min}}d\alpha\int^{\beta_{max}}_{\beta_{min}}d\beta}
\def\qq{\langle\bar qq\rangle}

\def\GGa{\langle GG\rangle}
\def\GGb{\langle g_s^2GG\rangle}

\def\qGqa{\langle\bar qGq\rangle}

\def\qGqb{\langle g_s\bar q\sigma Gq\rangle}

\def\FF(s){\left[(\alpha+\beta)m_c^2-\alpha\beta s\right]}
\def\HH(s){\left[m_c^2-\alpha(1-\alpha) s\right]}
\def\KK(s){\left[\gamma m_c^2-\gamma(1-\gamma) s\right]}
\def\non{\\ \nonumber}

\allowdisplaybreaks[3]

\begin{document}
\title{Charmonium-like states with the exotic quantum number $J^{PC} = 3^{-+}$}
%

\author{Hong-Zhou Xi$^1$}
\email{hzxi@seu.edu.cn}
\author{Hua-Xing Chen$^1$}
\email{hxchen@seu.edu.cn}
\author{Wei Chen$^2$}
\email{chenwei29@mail.sysu.edu.cn}
\author{T. G. Steele$^3$}
\email{tom.steele@usask.ca}
\author{Yong Zhang$^1$}
\email{zyfghe@seu.edu.cn}
\author{Dan Zhou$^4$}
\email{danzhou@hebtu.edu.cn}

\affiliation{$^1$School of Physics, Southeast University, Nanjing 210094, China\\
	$^2$School of Physics, Sun Yat-Sen University, Guangzhou 510275, China\\
	$^3$Department of Physics and Engineering Physics, University of Saskatchewan, Saskatoon, SK, S7N 5E2, Canada\\
	$^4$Department of Physics and Hebei Key Laboratory of Photophysics Research and Application, Hebei Normal University, Shijiazhuang 050024, China
}

\begin{abstract}
We apply the method of QCD sum rules to study the $q c \bar q \bar c$ tetraquark states with the exotic quantum number $J^{PC} = 3^{-+}$, and extract the mass of the lowest-lying state to be ${4.49^{+0.45}_{-0.41}}$~GeV. To construct the relevant tetraquark currents we need to explicitly add the covariant derivative operator. Our systematic analysis of these interpolating currents indicates that: a) this state readily decays into the $P$-wave $[\rho J/\psi] / [\omega J/\psi ]$ channel but not into the $ [\rho \chi_{c2}]/[\omega \chi_{c2}]/[J/\psi f_2(1270)]$ channels, and b) it readily decays into the $[D^* \bar D_2^*]$ channel but not into the $P$-wave $[D^* \bar D^*]$ channel.
\end{abstract}
\maketitle
\pagenumbering{arabic}
%
%
%
\section{Introduction}\label{sec:intro}
%

There have been many candidates for exotic hadrons observed in particle experiments, which can not be well explained in the traditional quark model~\cite{pdg,Liu:2019zoy,Lebed:2016hpi,Esposito:2016noz,Guo:2017jvc,Ali:2017jda,Olsen:2017bmm,Karliner:2017qhf,Brambilla:2019esw,Guo:2019twa}. Many of them still have the ``traditional'' quantum numbers that the traditional $\bar q q$ mesons and $qqq$ baryons can form. This makes them not so easy to be clearly identified as exotic hadrons. However, there exist some ``exotic'' quantum numbers that the traditional hadrons can not form, such as the spin-parity quantum numbers $J^{PC} = 0^{--}$, $0^{+-}$, $1^{-+}$, $2^{+-}$, $3^{-+}$, and $4^{+-}$, etc. These ``exotic'' quantum numbers are of particular interest, because the states with such quantum numbers can not be explained as traditional hadrons. Such states are definitely exotic hadrons, whose possible interpretations are tetraquark states~\cite{Chen:2008qw,Chen:2008ne,Jiao:2009ra,Huang:2016rro,LEE:2020eif,Du:2012pn,Fu:2018ngx,Dong:2022otb,Xi:2023byo,Dong:2022cuw,Yang:2022rck,Ji:2022blw,Wang:2021lkg,Wang:2023jaw}, hybrid states~\cite{Meyer:2015eta,Chetyrkin:2000tj,Zhang:2013rya,Huang:2014hya,Huang:2016upt,Ho:2018cat,Wang:2023whb,Su:2023jxb,Tan:2024grd,Chen:2022isv,Dudek:2013yja,Li:2021fwk,Tang:2021zti,Qiu:2022ktc,Wan:2022xkx,Wang:2022sib,Frere:2024wsf,Barsbay:2024vjt}, and glueballs~\cite{Qiao:2014vva,Tang:2015twt,Pimikov:2017bkk}, etc. Note that these exotic structures may mix together, making it challenging to arrive at a clear differentiation.

Among the above exotic quantum numbers, the hybrid states of $J^{PC} = 1^{-+}$ have been extensively studied in literature, since they are predicted to be the lightest hybrid states~\cite{Meyer:2015eta}, and there have been some experimental evidences on their existence~\cite{E852:1997gvf,CrystalBarrel:1999reg,E862:2006cfp}. The light tetraquark states of $J^{PC} = 1^{-+}$ have also been studied in Refs.~\cite{Chen:2008qw,Chen:2008ne} using the method of QCD sum rules, and their masses and possible decay channels were predicted there for both the isospin-0 and isospin-1 states. Later the same QCD sum rule method was applied to extensively study the light tetraquark states of $J^{PC} = 0^{--}/0^{+-}/2^{+-}/4^{+-}$ in Refs.~\cite{Jiao:2009ra,Huang:2016rro,LEE:2020eif,Du:2012pn,Fu:2018ngx,Dong:2022otb,Xi:2023byo}.

In this paper we shall study the exotic quantum number $J^{PC} = 3^{-+}$ using the method of QCD sum rules. The light $q s \bar q \bar s$  tetraquark states ($q=up/down$ and $s=strange$) with such a quantum number have been systematically investigated in Ref.~\cite{Su:2020reg}, and in this paper we shall further study their corresponding charmonium-like $q c \bar q \bar c$ tetraquark states. These states are potential exotic hadrons to be observed in the future BESIII, Belle-II, and LHCb experiments. There are just a few theoretical studies on this subject. In Ref.~\cite{Zhu:2013sca} the authors used the one-boson-exchange model to study the $D^* \bar D_2^*$ molecular state of $J^{PC} = 3^{-+}$, and their results suggest its possible existence. In Ref.~\cite{Dong:2021juy} the authors further investigated this state by solving the Bethe-Salpeter equation. Additionally, there was a Lattice QCD study on the $J^{PC} = 3^{-+}$ glueball~\cite{Shen:1984mxq}.

This paper is organized as follows. In Sec.~\ref{sec:current}, we systematically construct the $q c \bar q \bar c$ tetraquark currents with the exotic quantum number $J^{PC} = 3^{-+}$. Then we apply the QCD sum rule method to study them in Sec.~\ref{sec:sumrule}, and perform numerical analyses in Sec.~\ref{sec:numerical}. The obtained results are summarized and discussed in Sec.~\ref{sec:summary}.

%
\section{Interpolating Currents}\label{sec:current}
%

In this section we construct the hidden-charm tetraquark currents with the exotic quantum number $J^{PC} = 3^{-+}$. We have systematically constructed the hidden-strange tetraquark currents with such a quantum number in Ref.~\cite{Su:2020reg}, and in this paper we just need to replace the $strange$ quarks by the $charm$ quarks. Note that the exotic quantum number $J^{PC} = 3^{-+}$ can not be simply reached by using one quark field and one antiquark field, while it can not be reached by using two quark fields and two antiquark fields neither. Actually, we need two quark fields and two antiquark fields together with at least one derivative to reach such a quantum number.

As the first step, we work within the diquark-antidiquark configuration, where the derivative can be either inside the diquark/antidiquark field or between them ($c=charm$ and $q=up/down$):
\begin{eqnarray}
\eta &=& \big[c_a^T C \Gamma_1 {\overset{\leftrightarrow}{D}}_\alpha q_b \big] (\bar{c}_c \Gamma_2 C \bar{q}_d^T) \, ,
\label{eq:derivative1}
\\
\eta^\prime &=& (c_a^T C \Gamma_1 q_b) \big[\bar{c}_c \Gamma_2 C {\overset{\leftrightarrow}{D}}_\alpha \bar{q}_d^T\big] \, ,
\label{eq:derivative2}
\\
\eta^{\prime\prime} &=& \big[(c_a^T C \Gamma_3 q_b){\overset{\leftrightarrow}{D}}_\alpha(\bar{c}_c \Gamma_4 C \bar{q}_d^T)\big] \, .
\label{eq:derivative3}
\end{eqnarray}
Here $a \cdots d$ are color indices, and the sum over repeated indices is taken; $\Gamma_{1\cdots4}$ are Dirac matrices; $\big[ X {\overset{\leftrightarrow}{D}}_\alpha Y \big] = X [D_\alpha Y] - [D_\alpha X] Y$ with the covariant derivative $D_\alpha = \partial_\alpha + i g_s A_\alpha$. We find that only the former two can be combined to reach the exotic quantum number $J^{PC} = 3^{-+}$.

There are altogether six independent diquark-antidiquark currents of $J^{PC} = 3^{-+}$:
%
\begin{widetext}
\begin{eqnarray}
\eta^{1}_{\alpha_1\alpha_2\alpha_3} &=&
\epsilon^{abe} \epsilon^{cde} \times \mathcal{S} \Big\{
\big[c_a^T C \gamma_{\alpha_1} {\overset{\leftrightarrow}{D}}_{\alpha_3} q_b \big] (\bar{c}_c \gamma_{\alpha_2} C \bar{q}_d^T)
\label{def:eta1}
+ (c_a^T C \gamma_{\alpha_1} q_b)  \big[ \bar{c}_c \gamma_{\alpha_2} C {\overset{\leftrightarrow}{D}}_{\alpha_3} \bar{q}_d^T \big]
\Big\} \, ,
\\ \eta^{2}_{\alpha_1\alpha_2\alpha_3} &=&
(\delta^{ac} \delta^{bd} + \delta^{ad} \delta^{bc} ) \times \mathcal{S}\Big\{
\big[c_a^T C \gamma_{\alpha_1} {\overset{\leftrightarrow}{D}}_{\alpha_3} q_b \big] (\bar{c}_c \gamma_{\alpha_2} C \bar{q}_d^T)
\label{def:eta2}
+ (c_a^T C \gamma_{\alpha_1} q_b)  \big[ \bar{c}_c \gamma_{\alpha_2} C {\overset{\leftrightarrow}{D}}_{\alpha_3} \bar{q}_d^T \big]
\Big\} \, ,
\\ \eta^{3}_{\alpha_1\alpha_2\alpha_3} &=&
\epsilon^{abe} \epsilon^{cde} \times \mathcal{S}\Big\{
\big[c_a^T C \gamma_{\alpha_1}\gamma_5 {\overset{\leftrightarrow}{D}}_{\alpha_3} q_b \big] (\bar{c}_c \gamma_{\alpha_2}\gamma_5 C \bar{q}_d^T)
\label{def:eta3}
+ (c_a^T C \gamma_{\alpha_1}\gamma_5 q_b)  \big[ \bar{c}_c \gamma_{\alpha_2}\gamma_5 C {\overset{\leftrightarrow}{D}}_{\alpha_3} \bar{q}_d^T \big]
\Big\} \, ,
\\ \eta^{4}_{\alpha_1\alpha_2\alpha_3} &=&
(\delta^{ac} \delta^{bd} + \delta^{ad} \delta^{bc} ) \times \mathcal{S} \Big\{
\big[c_a^T C \gamma_{\alpha_1}\gamma_5 {\overset{\leftrightarrow}{D}}_{\alpha_3} q_b \big] (\bar{c}_c \gamma_{\alpha_2}\gamma_5 C \bar{q}_d^T)
\label{def:eta4}
+ (c_a^T C \gamma_{\alpha_1}\gamma_5 q_b)  \big[ \bar{c}_c \gamma_{\alpha_2}\gamma_5 C {\overset{\leftrightarrow}{D}}_{\alpha_3} \bar{q}_d^T \big]
\Big\} \, ,
\\ \eta^{5}_{\alpha_1\alpha_2\alpha_3} &=&
\epsilon^{abe} \epsilon^{cde} \times g^{\mu\nu} \mathcal{S}\Big\{
\big[c_a^T C \sigma_{\alpha_1\mu} {\overset{\leftrightarrow}{D}}_{\alpha_3} q_b \big] (\bar{c}_c \sigma_{\alpha_2\nu} C \bar{q}_d^T)
\label{def:eta5}
+ (c_a^T C \sigma_{\alpha_1\mu} q_b)  \big[ \bar{c}_c \sigma_{\alpha_2\nu} C {\overset{\leftrightarrow}{D}}_{\alpha_3} \bar{q}_d^T \big]
\Big\} \, ,
\\ \eta^{6}_{\alpha_1\alpha_2\alpha_3} &=&
(\delta^{ac} \delta^{bd} + \delta^{ad} \delta^{bc} ) \times g^{\mu\nu} \mathcal{S}\Big\{
\big[c_a^T C \sigma_{\alpha_1\mu} {\overset{\leftrightarrow}{D}}_{\alpha_3} q_b \big] (\bar{c}_c \sigma_{\alpha_2\nu} C \bar{q}_d^T)
\label{def:eta6}
+ (c_a^T C \sigma_{\alpha_1\mu} q_b)  \big[ \bar{c}_c \sigma_{\alpha_2\nu} C {\overset{\leftrightarrow}{D}}_{\alpha_3} \bar{q}_d^T \big]
\Big\} \, ,
\end{eqnarray}
\end{widetext}
%
where $\mathcal{S}$ denotes symmetrization and subtracting the trace terms in the set $\{\alpha_1\alpha_2\alpha_3\}$, so that the spin-3 components can be well separated. Three of them $\eta^{1,3,5}_{\alpha_1\alpha_2\alpha_3}$ have the antisymmetric color structure $(qc)_{\mathbf{\bar 3}_C}(\bar q \bar c)_{\mathbf{3}_C}$, and the other three $\eta^{2,4,6}_{\alpha_1\alpha_2\alpha_3}$ have the symmetric color structure $(qc)_{\mathbf{6}_C}(\bar q \bar c)_{\mathbf{\bar 6}_C}$.

Besides the diquark-antidiquark configuration, we also investigate the meson-meson configuration. There are six independent meson-meson currents of $J^{PC} = 3^{-+}$:
%
\begin{eqnarray}
\xi^{1}_{\alpha_1\alpha_2\alpha_3} &=&
\label{def:xi1}
\mathcal{S}\Big\{
(\bar{c}_a \gamma_{\alpha_1} c_a) {\overset{\leftrightarrow}{D}}_{\alpha_3}(\bar{q}_b \gamma_{\alpha_2} q_b) \Big\} \, ,
\\ \xi^{2}_{\alpha_1\alpha_2\alpha_3} &=&
\label{def:xi2}
\mathcal{S}\Big\{
(\bar{c}_a \gamma_{\alpha_1} \gamma_5 c_a) {\overset{\leftrightarrow}{D}}_{\alpha_3}(\bar{q}_b \gamma_{\alpha_2} \gamma_5 q_b) \Big\} \, ,
\\ \xi^{3}_{\alpha_1\alpha_2\alpha_3} &=&
g^{\mu\nu} \mathcal{S}\Big\{
(\bar{c}_a \sigma_{\alpha_1\mu} c_a) {\overset{\leftrightarrow}{D}}_{\alpha_3}(\bar{q}_b \sigma_{\alpha_2\nu} q_b)
\label{def:xi3}
\Big\} \, ,
\\ \xi^{4}_{\alpha_1\alpha_2\alpha_3} &=&
\mathcal{S}\Big\{
\big[ \bar{c}_a \gamma_{\alpha_1} {\overset{\leftrightarrow}{D}}_{\alpha_3} q_a \big] (\bar{q}_b \gamma_{\alpha_2} c_b)
\label{def:xi4}
\\ \nonumber && ~~~~~~~~~~~ - (\bar{c}_a \gamma_{\alpha_1} q_a) \big [\bar{q}_b \gamma_{\alpha_2} {\overset{\leftrightarrow}{D}}_{\alpha_3} c_b \big]
\Big\} \, ,
\\ \xi^{5}_{\alpha_1\alpha_2\alpha_3} &=&
\mathcal{S}\Big\{
\big[ \bar{c}_a \gamma_{\alpha_1} \gamma_5 {\overset{\leftrightarrow}{D}}_{\alpha_3} q_a \big] (\bar{q}_b \gamma_{\alpha_2} \gamma_5 c_b)
\label{def:xi5}
\\ \nonumber && ~~~~~~~~~~~ - (\bar{c}_a \gamma_{\alpha_1} \gamma_5 q_a) \big [\bar{q}_b \gamma_{\alpha_2} \gamma_5 {\overset{\leftrightarrow}{D}}_{\alpha_3} c_b \big]
\Big\} \, ,
\\ \xi^{6}_{\alpha_1\alpha_2\alpha_3} &=&
g^{\mu\nu} \mathcal{S}\Big\{
\big[ \bar{c}_a \sigma_{\alpha_1\mu} {\overset{\leftrightarrow}{D}}_{\alpha_3} q_a \big] (\bar{q}_b \sigma_{\alpha_2\nu} c_b)
\label{def:xi6}
\\ \nonumber && ~~~~~~~~~~~ - (\bar{c}_a \sigma_{\alpha_1\mu} q_a) \big [\bar{q}_b \sigma_{\alpha_2\nu} {\overset{\leftrightarrow}{D}}_{\alpha_3} c_b \big]
\Big\} \, .
\end{eqnarray}
%
We can apply the Fierz rearrangement to relate the above diquark-antidiquark and meson-meson currents:
%
\begin{eqnarray}
\label{eq:fierz}
&& \left(\begin{array}{c}
\eta^{1}_{\alpha_1\alpha_2\alpha_3}
\\
\eta^{2}_{\alpha_1\alpha_2\alpha_3}
\\
\eta^{3}_{\alpha_1\alpha_2\alpha_3}
\\
\eta^{4}_{\alpha_1\alpha_2\alpha_3}
\\
\eta^{5}_{\alpha_1\alpha_2\alpha_3}
\\
\eta^{6}_{\alpha_1\alpha_2\alpha_3}
\end{array}\right)
=
\\ \nonumber &&
\left(\begin{array}{cccccc}
-{1\over2} &  {1\over2} &  {1\over2} & -{1\over2} &  {1\over2} &  {1\over2}
\\
-{1\over2} &  {1\over2} &  {1\over2} &  {1\over2} & -{1\over2} & -{1\over2}
\\
 {1\over2} & -{1\over2} &  {1\over2} & -{1\over2} &  {1\over2} & -{1\over2}
\\
 {1\over2} & -{1\over2} &  {1\over2} &  {1\over2} & -{1\over2} &  {1\over2}
\\
 1         & 1          &  0         &  1         &  1         & 0
\\
 1         & 1          &  0         & -1         & -1         & 0
\end{array}\right)
\left(\begin{array}{c}
\xi^{1}_{\alpha_1\alpha_2\alpha_3}
\\
\xi^{2}_{\alpha_1\alpha_2\alpha_3}
\\
\xi^{3}_{\alpha_1\alpha_2\alpha_3}
\\
\xi^{4}_{\alpha_1\alpha_2\alpha_3}
\\
\xi^{5}_{\alpha_1\alpha_2\alpha_3}
\\
\xi^{6}_{\alpha_1\alpha_2\alpha_3}
\end{array}\right) \, .
\end{eqnarray}
%
Hence, these two configurations are equivalent, and we shall apply this Fierz identity to study the decay properties at the end of this paper. However, this equivalence is just between the local diquark-antidiquark and meson-meson currents, while the tightly-bound diquark-antidiquark tetraquark states and the weakly-bound meson-meson molecular states are totally different. To clearly describe them, we need to investigate the non-local currents, which can not be done within the QCD sum rule framework yet.

%
\section{QCD sum rule Analysis}
\label{sec:sumrule}
%

In this section we apply the QCD sum rule method to study the six diquark-antidiquark currents $\eta^i_{\alpha_1\alpha_2\alpha_3}$ ($i=1 \cdots 6$), and calculate their two-point correlation functions
%
\begin{eqnarray}
&& \Pi^{ii}_{\alpha_1\alpha_2\alpha_3,\beta_1\beta_2\beta_3}(q^2)
\label{def:pi}
\\ \nonumber &\equiv& i \int d^4x e^{iqx} \langle 0 | {\bf T}[ \eta^i_{\alpha_1\alpha_2\alpha_3}(x) \eta^{i,\dagger}_{\beta_1\beta_2\beta_3} (0)] | 0 \rangle
\\ \nonumber &=& (-1)^J~\mathcal{S}^\prime [\tilde g_{\alpha_1 \beta_1} \tilde g_{\alpha_2 \beta_2} \tilde g_{\alpha_3 \beta_3}]~\Pi_{ii} (q^2) \, ,
\end{eqnarray}
%
at both the hadron and quark-gluon levels. Here $\tilde g_{\mu \nu} = g_{\mu \nu} - q_\mu q_\nu / q^2$, and $\mathcal{S}^\prime$ denotes symmetrization and subtracting trace terms in the two sets $\{\alpha_1\alpha_2\alpha_3\}$ and $\{\beta_1\beta_2\beta_3\}$.

Take the first current $\eta^1_{\alpha_1\alpha_2\alpha_3}$ as an example. We assume that it couples to the possibly-existing exotic state $X_1$ through
\begin{equation}
\label{eq:defg}
\langle 0| \eta^1_{\alpha_1\alpha_2\alpha_3} | X_1 \rangle = f_1 \epsilon_{\alpha_1\alpha_2\alpha_3} \, ,
\end{equation}
with $f_1$ the decay constant. The symmetric and traceless polarization tensor $\epsilon_{\alpha_1\alpha_2\alpha_3}$ satisfies
\begin{equation}
\sum_{spin}\epsilon_{\alpha_1\alpha_2\alpha_3} \epsilon^*_{\beta_1\beta_2\beta_3} = \mathcal{S}^\prime [\tilde g_{\alpha_1 \beta_1} \tilde g_{\alpha_2 \beta_2} \tilde g_{\alpha_3 \beta_3}] \, .
\end{equation}

At the hadron level we apply the dispersion relation to write Eq.~(\ref{def:pi}) as:
%
\begin{equation}
\Pi_{11}(q^2) = \int^\infty_{4 m_c^2}\frac{\rho^{\rm phen}_{11}(s)}{s-q^2-i\varepsilon}ds \, ,
\label{eq:hadron}
\end{equation}
%
with $\rho^{\rm phen}_{11}(s)$ the phenomenological spectral density. We parameterize it using one pole dominance for the state $X_1$ and a continuum contribution
%
\begin{eqnarray}
&& \rho^{\rm phen}_{11}(s) \times \mathcal{S}^\prime [\cdots]
\\ \nonumber &\equiv& \sum_n\delta(s-M^2_n) \langle 0| \eta^1_{\alpha_1\alpha_2\alpha_3} | n \rangle \langle n | \eta^{1,\dagger}_{\beta_1\beta_2\beta_3} |0 \rangle
\\ \nonumber &=& f^2_1 \delta(s-M^2_1) \times \mathcal{S}^\prime [\cdots] + \rm{continuum} \, .
\label{eq:rho}
\end{eqnarray}
%

At the quark-gluon level we insert the current $\eta^1_{\alpha_1\alpha_2\alpha_3}$ into Eq.~(\ref{def:pi}), and calculate it using the method of operator product expansion (OPE), from which we extract the OPE spectral density $\rho_{11}(s) \equiv \rho_{11}^{\rm OPE}(s)$. Then we perform the Borel transformation at both the hadron and quark-gluon levels. After approximating the continuum using the OPE spectral density above the threshold value $s_0$, we obtain the QCD sum rule equation
%
\begin{equation}
\Pi_{11}(s_0, M_B^2) \equiv f^2_1 e^{-M_1^2/M_B^2} = \int^{s_0}_{4 m_c^2} e^{-s/M_B^2}\rho_{11}(s)ds \, .
\label{eq_fin}
\end{equation}
%
We can use it to further calculate $M_1$ through
%
\begin{eqnarray}
M_{1}^2(s_0, M_B) &=& \frac{\frac{\partial}{\partial(-1/M_B^2)}\Pi_{11}(s_0, M_B^2)}{\Pi_{11}(s_0, M_B^2)}
\label{eq:LSR}
\\ \nonumber &=& \frac{\int^{s_0}_{4 m_c^2} e^{-s/M_B^2}s\rho_{11}(s)ds}{\int^{s_0}_{4 m_c^2} e^{-s/M_B^2}\rho_{11}(s)ds} \, .
\end{eqnarray}
%
\begin{widetext}
    The OPE spectral density $\rho_{11}(s)$ extracted from the current $\eta^1_{\alpha_1\alpha_2\alpha_3}$ is
	\begin{equation}
		\rho_{11}(s) = \rho^{pert}_{11}(s) + \rho^{\qq}_{11}(s) + \rho^{\GGa}_{11}(s)+ \rho^{\qGqa}_{11}(s) + \rho^{\qq^2}_{11}(s)  + \rho^{\qq\qGqa}_{11}(s)+\rho^{\qGqa^2}_{11}(s) \, ,
		\label{ope:eta1}
	\end{equation}
	where
	\begin{eqnarray}
		\nonumber \rho^{pert}_{11}(s) &=& \dab \Bigg\{ -\mathcal{F}(s)^5 \times \frac{1}{15360 \pi ^6 \alpha ^4 \beta ^4}  \times ( 3 \alpha ^5+33 \alpha ^4 \beta -4 \alpha ^4+102 \alpha ^3 \beta ^2-38 \alpha ^3 \beta
		\non
		&+&138 \alpha ^2 \beta ^3-90 \alpha ^2 \beta ^2+87 \alpha  \beta ^4-82 \alpha  \beta ^3+\alpha +21 \beta ^5-26 \beta ^4+5 \beta ) \Bigg\} \, ,
		\non
		\rho^{\qq}_{11}(s) &=& {\qq } \dab \Bigg\{ m_c \mathcal{F}(s)^3  \times  \frac{-6 \alpha ^3-24 \alpha ^2 \beta +6 \alpha ^2-26 \alpha  \beta ^2+15 \alpha  \beta -8 \beta ^3+7 \beta ^2}{96 \pi ^4 \alpha ^3 \beta ^2}  \Bigg\} \, ,
		\non
		\rho^{\GGa}_{11}(s) &=& {\GGb }  \int^{\alpha_{max}}_{\alpha_{min}}d\alpha  \Bigg\{ \int^{\beta_{max}}_{\beta_{min}}d\beta \Bigg\{\mathcal{F}(s)^2 \times \frac{1}{221184 \pi^6 \alpha^3 \beta^4} \times ( \mathcal{F}(s) \times (36 \alpha ^5 \beta +216 \alpha ^4 \beta ^2-127 \alpha ^4 \beta
		\non
		&+&612 \alpha ^3 \beta ^3 -570 \alpha ^3 \beta ^2+82 \alpha ^3 \beta +828 \alpha ^2 \beta ^4-1170 \alpha ^2 \beta ^3+237 \alpha ^2 \beta ^2+504 \alpha  \beta ^5-1006 \alpha  \beta ^4
		\non
		&+&417 \alpha  \beta ^3+10 \alpha \beta +108 \beta ^6-279 \beta ^5+142 \beta ^4+12 \beta ^2-\beta) + 24m_c^2 (-12 \alpha ^7-60 \alpha ^6 \beta +15 \alpha ^6
		\non
		&-&120 \alpha ^5 \beta ^2+60 \alpha ^5 \beta -120 \alpha ^4 \beta ^3+90 \alpha ^4 \beta ^2-60 \alpha ^3 \beta ^4 +60 \alpha ^3 \beta ^3-3 \alpha ^3-12 \alpha ^2 \beta ^5+15 \alpha ^2 \beta ^4-3 \alpha ^2 \beta)  ) \Bigg\}
		\non
		&+&\mathcal{H}(s)^3 \times \frac{1}{36864 \pi^6 (\alpha -1) \alpha^2}\Bigg\} \, ,
		\non
		\rho^{\qGqa}_{11}(s) &=& {\qGqb } \int^{\alpha_{max}}_{\alpha_{min}}d\alpha  \Bigg\{ \int^{\beta_{max}}_{\beta_{min}}d\beta \Bigg\{
		m_c \mathcal{F}(s)^2 \times  \frac{6 \alpha ^3-60 \alpha ^2 \beta -6 \alpha ^2-216 \alpha  \beta ^2+27 \alpha  \beta -98 \beta ^3+33 \beta ^2}{1536 \pi ^4 \alpha ^2 \beta ^2}  \Bigg\}
		\non
		&+& m_c \mathcal{H}(s)^2 \times  \frac{26 \alpha +13}{1536 \pi ^4 \alpha ^2}  \Bigg\} \, ,
		\non
		\rho^{\qq^2}_{11}(s)&=& {\qq^2 } \int^{\alpha_{max}}_{\alpha_{min}}d\alpha \Bigg\{ m_c^2 \mathcal{H}(s) \times  \frac{1}{12 \pi ^2 \alpha}  \Bigg\} \, ,
		\non
		\rho^{\qq\qGqa}_{11}(s)&=& {\qq\qGqb } \Bigg\{ \int^{\alpha_{max}}_{\alpha_{min}}d\alpha \Bigg\{  \frac{m_c^2 (118 \alpha ^2-131 \alpha -1)}{576 \pi ^2 \alpha}  \Bigg\}  + \int^{1}_{0} \delta\left( s - {m_c^2 \over (1-\alpha) \alpha} \right) d\alpha  \Bigg\{ -\frac{m_c^4}{24 \pi ^2 \alpha}  \Bigg\}  \Bigg\} \, ,
		\non
		\rho^{\qGqa^2}_{11}(s)&=&{\qGqb^2 } \int^{1}_{0} \delta\left( s - {m_c^2 \over (1-\alpha) \alpha} \right) d\alpha  \Bigg\{ \frac{m_c^2 (164 \alpha ^3-354 \alpha ^2+187 \alpha +4)}{2304 \pi ^2 (\alpha -1) \alpha} + \frac{m_c^4 (-94 \alpha ^2+107 \alpha +1)}{2304 \pi ^2 (\alpha -1) \alpha ^2 M_B^2}
		\non
		&+& \frac{m_c^6} {192 \pi ^2 (\alpha -1) \alpha ^2 M_B^4} \Bigg\} \, .
	\end{eqnarray}
\end{widetext}
In the above expressions, $\mathcal{F}(s) = \left[(\alpha + \beta) m_c^2 - \alpha \beta s\right]$, $\mathcal{H}(s) = \left[m_c^2 - \alpha(1 - \alpha)s\right]$, $\alpha_{\min} = \frac{1 - \sqrt{1 - 4m_c^2 / s}}{2}$, $\alpha_{\max} = \frac{1 + \sqrt{1 - 4m_c^2 / s}}{2}$, $\beta_{\min} = \frac{\alpha m_c^2}{\alpha s - m_c^2}$, and $\beta_{\max} = 1 - \alpha$. We have calculated the QCD spectral density $\rho_{11}(s)$ at the leading order of $\alpha_s$ and up to the dimension ten ($D=10$). In the calculations we have considered the perturbative term, the charm quark mass, the quark condensate $\langle \bar q q \rangle$, the gluon condensate $\langle g_s^2 GG \rangle$, the quark-gluon mixed condensate $\langle g_s \bar q \sigma G q \rangle$, and their combinations. We have ignored the chirally suppressed terms with light quark masses, and we have adopted the factorization assumption of vacuum saturation for higher dimensional condensates. We find that the $D=3$ term $\langle \bar q q \rangle$ and the $D=5$ term $\langle g_s \bar q \sigma G q \rangle$ are both multiplied by the charm quark mass, so they are important power corrections to the correlation functions. The QCD sum rule results extracted from the other five currents $\eta^{2\cdots6}_{\alpha_1\alpha_2\alpha_3}$ are given in Appendix~\ref{app:ope}. Based on these results, we shall perform numerical analyses in the next section.

\section{Numerical Analyses}\label{sec:numerical}
%
In this section we use the spectral densities given in Eq.~(\ref{ope:eta1}) and Eqs.(\ref{ope:eta2}-\ref{ope:eta6}) to perform numerical analyses. We shall use the following values for various QCD sum rule parameters~\cite{Yang:1993bp,Narison:2002hk,Gimenez:2005nt,Jamin:2002ev,Ioffe:2002be,Ovchinnikov:1988gk,Ellis:1996xc,Narison:2011xe,Narison:2018dcr,pdg}:
%
\begin{eqnarray}
\nonumber  \langle g_s^2 GG\rangle &=& 4 \pi \times (0.0635\pm0.0035) \mbox{ GeV}^4 \, ,
\\ \nonumber  \langle\bar qq \rangle &=& -(0.240 \pm 0.010)^3 \mbox{ GeV}^3 \, ,
\\ \langle g_s\bar q\sigma G q\rangle &=& - M_0^2\times\langle\bar qq\rangle \, ,
\label{condensates}
\\ \nonumber M_0^2 &=& (0.8 \pm 0.2) \mbox{ GeV}^2 \, ,
\\ \nonumber m_c(m_c) &=& 1.275 ^{+0.025}_{-0.035} \mbox{ GeV} \, .
\end{eqnarray}
%
The gluon condensate $\langle g_s^2 GG\rangle$ is still not well known, and the above value for this condensate is taken from Ref.~\cite{Narison:2018dcr}, which was updated in 2018. We note that this condensate does not contribute much to the spectral densities. Different with some other QCD sum rule calculations~\cite{Su:2023jxb,Tan:2024grd}, there is a minus sign in the definition of the mixed condensate $\langle g_s\bar q\sigma G q\rangle$, which is just because the definition of coupling constant $g_s$ is different~\cite{Yang:1993bp,Hwang:1994vp}, {\it i.e.}, $D_\alpha = \partial_\alpha + i g_s A_\alpha$ is used the present study, while $D_\alpha = \partial_\alpha - i g_s A_\alpha$ is used in Refs.~\cite{Su:2023jxb,Tan:2024grd}.

We take the spectral density $\rho_{11}(s)$ extracted from the current $\eta^1_{\alpha_1\alpha_2\alpha_3}$ as an example. As shown in Eq.~(\ref{eq:LSR}), the mass $M_X$ and the decay constant $f_{X}$ both depend on two free parameters: the threshold value $s_0$ and the Borel mass $M_B$. We investigate three aspects to find their proper working regions: a) the convergence of OPE, b) the one-pole-dominance assumption, and c) the mass dependence and the decay constant dependence on these two parameters.

%
\begin{figure}[hbt]
\begin{center}
\includegraphics[width=0.47\textwidth]{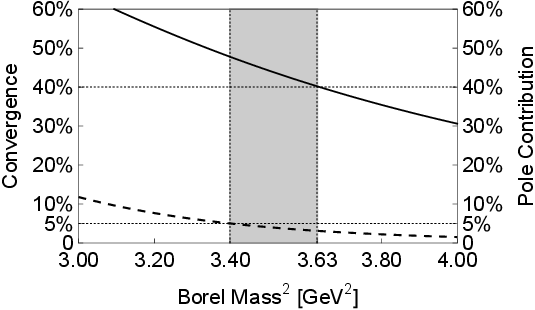}
\caption{CVG and PC as functions of the Borel mass $M_B$. These curves are extracted from the current $\eta^1_{\alpha_1\alpha_2\alpha_3}$ when setting $s_0 = 30.0$~GeV$^2$.}
\label{fig:cvgpole}
\end{center}
\end{figure}

Firstly, we investigate the convergence of OPE and require the $D=10$ terms to be less than 5\%:
\begin{equation}
\mbox{CVG} \equiv \left|\frac{ \Pi_{11}^{D=10}(\infty, M_B^2) }{ \Pi_{11}(\infty, M_B^2) }\right| \leq 5\% \, .
\end{equation}
As shown in Fig.~\ref{fig:cvgpole}, we determine the lower bound of the Borel mass to be $M_B^2 \geq 3.40$~GeV$^2$.

Secondly, we investigate the one-pole-dominance assumption and require the pole contribution (PC) to be larger than 40\%:
\begin{equation}
\mbox{PC} \equiv \left|\frac{ \Pi_{11}(s_0, M_B^2) }{ \Pi_{11}(\infty, M_B^2) }\right| \geq 40\% \, .
\label{eq:pole}
\end{equation}
As shown in Fig.~\ref{fig:cvgpole}, we determine the upper bound of the Borel mass to be $M_B^2 \leq 3.63$~GeV$^2$ when setting $s_0 = 30.0$~GeV$^2$. Altogether we determine the Borel window to be $3.40$~GeV$^2 \leq M_B^2 \leq 3.63$~GeV$^2$ when setting $s_0 = 30.0$~GeV$^2$. After changing $s_0$ and redoing the same procedures, we find that there are non-vanishing Borel windows as long as $s_0 \geq s_0^{\rm min} = 28.4$~GeV$^2$.

\begin{figure*}[]
\begin{center}
\includegraphics[width=0.45\textwidth]{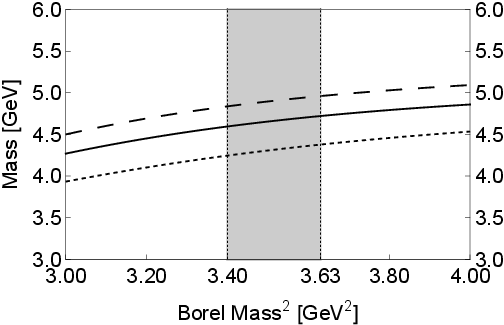}
~~~~~
\includegraphics[width=0.45\textwidth]{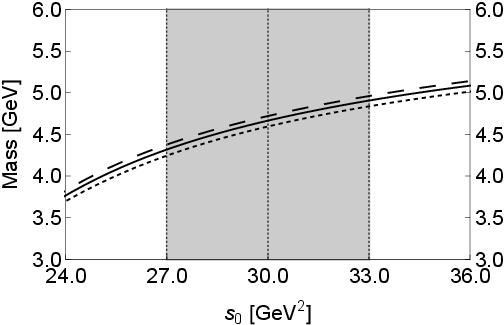}
\caption{
Mass calculated using the current $\eta^{1}_{\alpha_1\alpha_2\alpha_3}$ with respect to the Borel mass $M_B$ (left) and the threshold value $s_0$ (right).
In the left panel the short-dashed/solid/long-dashed curves are obtained by setting $s_0 = 27.0/30.0/33.0$ GeV$^2$, respectively.
In the right panel the short-dashed/solid/long-dashed curves are obtained by setting $M_B^2 = 3.40/3.52/3.63$ GeV$^2$, respectively.}
\label{fig:eta1mass}
\end{center}
\end{figure*}

\begin{figure*}[]
\begin{center}
\includegraphics[width=0.45\textwidth]{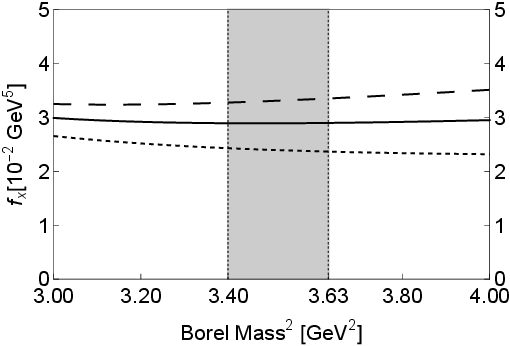}
~~~~~
\includegraphics[width=0.45\textwidth]{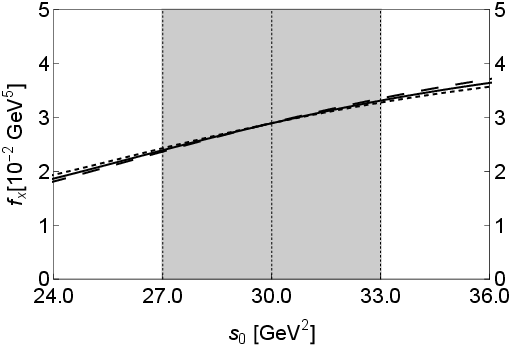}
\caption{
Decay constant calculated using the current $\eta^{1}_{\alpha_1\alpha_2\alpha_3}$ with respect to the Borel mass $M_B$ (left) and the threshold value $s_0$ (right). In the left panel the short-dashed/solid/long-dashed curves are obtained by setting $s_0 = 27.0/30.0/33.0$ GeV$^2$, respectively.
In the right panel the short-dashed/solid/long-dashed curves are obtained by setting $M_B^2 = 3.40/3.52/3.63$ GeV$^2$, respectively.}
\label{fig:eta1fx}
\end{center}
\end{figure*}

Thirdly, we investigate the mass dependence and the decay constant dependence on $s_0$ and $M_B$. We respectively show the mass $M_1$ in Fig.~\ref{fig:eta1mass} and the decay constant $f_1$ in Fig.~\ref{fig:eta1fx} as functions of these two parameters. Their dependence on $M_B$ is weak inside the Borel window $3.40$~GeV$^2 \leq M_B^2 \leq 3.63$~GeV$^2$, and their dependence on $s_0$ is moderate and acceptable around $s_0 \sim 30.0$~GeV$^2$. Accordingly, we choose our working regions to be $27.0$~GeV$^2 \leq s_0 \leq 33.0$~GeV$^2$ and $3.40$~GeV$^2 \leq M_B^2 \leq 3.63$~GeV$^2$, where the mass $M_1$ is evaluated to be
\begin{equation}
M_1 = {4.66^{+0.49}_{-0.46}}{\rm~GeV} \, .
\label{eq:mass1}
\end{equation}
Its central value corresponds to $s_0=30.0$~GeV$^2$ and $M_B^2 = 3.52$~GeV$^2$. Its uncertainty is due to $s_0$, $M_B$, and various QCD sum rule parameters listed in Eqs.~(\ref{condensates}).

We apply the same procedures to study the other five currents $\eta^{2\cdots6}_{\alpha_1\alpha_2\alpha_3}$, and summarize their results in Table~\ref{tab:results}. Especially, the mass $M_2$ extracted from the current $\eta^{2}_{\alpha_1\alpha_2\alpha_3}$ is calculated to be
\begin{equation}
M_2 = {4.50^{+0.45}_{-0.41}}{\rm~GeV} \, ,
\label{eq:mass2}
\end{equation}
which is slightly smaller than the mass $M_1$ extracted from the current $\eta^{1}_{\alpha_1\alpha_2\alpha_3}$, while the masses extracted from the other four currents $\eta^{3\cdots6}_{\alpha_1\alpha_2\alpha_3}$ are all significantly larger.

\begin{figure}[hbt]
\begin{center}
\includegraphics[width=0.47\textwidth]{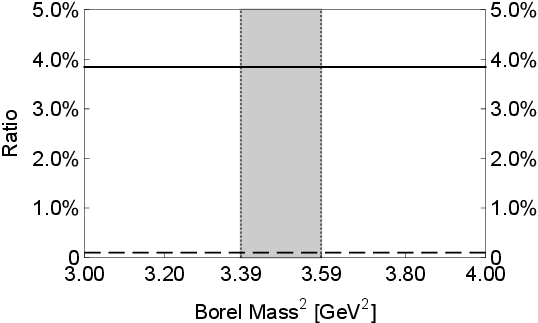}
\end{center}
\caption{The normalized off-diagonal correlation functions $\left|\Pi_{12}/\sqrt{\Pi_{11}\Pi_{22}}\right|$ (solid) and $\left|\Pi^\prime_{12}/\sqrt{\Pi^\prime_{11}\Pi^\prime_{22}}\right|$ (dashed) as functions of the Borel mass $M_B$. These curves are obtained using the two currents $\eta^{1,2}_{\alpha_1\alpha_2\alpha_3}$ and their mixing currents $J^{1,2}_{\alpha_1\alpha_2\alpha_3}$  by setting  $s_0 = 29.0$~GeV$^2$.}
\label{fig:offdiagonal}
\end{figure}

It is interesting to investigate the mixing of $\eta^{1}_{\alpha_1\alpha_2\alpha_3}$ and $\eta^{2}_{\alpha_1\alpha_2\alpha_3}$ by calculating their off-diagonal correlation function, {\it i.e.}, the ``12'' component of Eq.~(\ref{def:pi}):
\begin{eqnarray}
&& \Pi^{12}_{\alpha_1\alpha_2\alpha_3,\beta_1\beta_2\beta_3}(q^2)
\\ \nonumber &\equiv& i \int d^4x e^{iqx} \langle 0 | {\bf T}[ \eta^1_{\alpha_1\alpha_2\alpha_3}(x) \eta^{2,\dagger}_{\beta_1\beta_2\beta_3} (0)] | 0 \rangle
\\ \nonumber &=& (-1)^J~\mathcal{S}^\prime [\tilde g_{\alpha_1 \beta_1} \tilde g_{\alpha_2 \beta_2} \tilde g_{\alpha_3 \beta_3}]~\Pi_{12} (q^2)  \, .
\end{eqnarray}
To see how large it is, we choose $s_0 = 29.0$ GeV$^2$ and $M_B^2 = 3.40 $ GeV$^2$ to obtain
\begin{equation}
\left(
\begin{array}{cc}
\Pi_{11} & \Pi_{12}  \\
\Pi_{21} & \Pi_{22}  \\
\end{array}
\right)
=
\left(
\begin{array}{cc}
70.27 & 3.88  \\
3.88  & 145.14  \\
\end{array}
\right) \times 10^{-5} {\rm~GeV}^{10} ,
\label{eq:mixingpi1}
\end{equation}
which indicates that $\eta^1_{\alpha_1\alpha_2\alpha_3}$ and $\eta^2_{\alpha_1\alpha_2\alpha_3}$ are weakly correlated with each other, as shown in Fig.~\ref{fig:offdiagonal}.

To diagonalize the $2\times2$ matrix given in Eq.~(\ref{eq:mixingpi1}), we construct two mixing currents $J^{1,2}_{\alpha_1\alpha_2\alpha_3}$:
\begin{equation}
\left(\begin{array}{c}
J^1_{\alpha_1\alpha_2\alpha_3}
\\
J^2_{\alpha_1\alpha_2\alpha_3}
\end{array}\right)
=
\mathbb{T}
\left(\begin{array}{c}
\eta^1_{\alpha_1\alpha_2\alpha_3}
\\
\eta^2_{\alpha_1\alpha_2\alpha_3}
\end{array}\right)
\, ,
\label{eq:transition}
\end{equation}
where $\mathbb{T}$ is defined as the transition matrix. We  apply the method of operator product expansion to calculate the two-point correlation functions of the mixing currents $J^{1,2}_{\alpha_1\alpha_2\alpha_3}$ ($i,j=1,2$):
\begin{eqnarray}
&& \Pi^{\prime,ij}_{\alpha_1\alpha_2\alpha_3,\beta_1\beta_2\beta_3}(q^2)
\\ \nonumber &\equiv& i \int d^4x e^{iqx} \langle 0 | {\bf T}[ J^i_{\alpha_1\alpha_2\alpha_3}(x) J^{j,\dagger}_{\beta_1\beta_2\beta_3} (0)] | 0 \rangle
\\ \nonumber &=& (-1)^J~\mathcal{S}^\prime [\tilde g_{\alpha_1 \beta_1} \tilde g_{\alpha_2 \beta_2} \tilde g_{\alpha_3 \beta_3}]~\Pi^\prime_{ij} (q^2)  \, .
\end{eqnarray}
After choosing
\begin{equation}
\mathbb{T}
=
\left(
\begin{array}{cc}
 1.00 & -0.05 \\
 0.05 & 1.00 \\
\end{array}
\right)\, ,
\label{eq:tmatrix}
\end{equation}
we obtain
\begin{equation}
\left(
\begin{array}{cc}
\Pi^\prime_{11} & \Pi^\prime_{12}  \\
\Pi^\prime_{21} & \Pi^\prime_{22}  \\
\end{array}
\right)
=
\left(
\begin{array}{cc}
145.34 & 0 \\
0 & 70.07 \\
\end{array}
\right)
 \times 10^{-5} {\rm~GeV}^{10} ,
\label{eq:mixingpi2}
\end{equation}
at $s_0 = 29.0$~GeV$^2$ and $M_B^2 = 3.40$~GeV$^2$. Consequently, the off-diagonal terms of $J^{1}_{\alpha_1\alpha_2\alpha_3}$ and $J^{2}_{\alpha_1\alpha_2\alpha_3}$ are negligible, indicating that these two mixing currents are non-correlated around here, as shown in Fig.~\ref{fig:offdiagonal}. Implicitly, the above mixing analysis can work because the continuum is basically the same in both the $\eta^{1}_{\alpha_1\alpha_2\alpha_3}$ and $\eta^{2}_{\alpha_1\alpha_2\alpha_3}$ channels, which allows us to simplify the analysis by choosing a continuum for the off-diagonal correlation function.

We apply the same procedures to study $J^{1,2}_{\alpha_1\alpha_2\alpha_3}$, and the obtained results are summarized in Table~\ref{tab:results}. Especially, the mass extracted from the mixing current $J^2_{\alpha_1\alpha_2\alpha_3}$ is slightly reduced from the single current $\eta^2_{\alpha_1\alpha_2\alpha_3}$ to be
\begin{equation}
M_2^\prime = {4.49^{+0.45}_{-0.41}} {\rm~GeV} \, .
\end{equation}
%
\section{Summary and Discussions}
\label{sec:summary}
%

\begin{table*}[hpt]
\begin{center}
\renewcommand{\arraystretch}{1.8}
\caption{QCD sum rule results extracted from the diquark-antidiquark currents $\eta^{1\cdots6}_{\alpha_1\alpha_2\alpha_3}$ and the mixing currents $J^{1\cdots2}_{\alpha_1\alpha_2\alpha_3}$ with the exotic quantum number $J^{PC} = 3^{-+}$.}
\begin{tabular}{c|c|c|c|c|c }
\hline\hline
~Currents~&~~$M_B^2~[{\rm GeV}^2]$~~&~$s_0~[{\rm GeV}^2]$~&~~Pole~[\%]~~&~~Mass~[GeV]~~&~$f_X~[{\rm GeV}^5]$
\\ \hline \hline
$\eta^1_{\alpha_1\alpha_2\alpha_3}$&$3.40$-$3.63$&$30.0\pm3.0$&$40$-$48$&${4.66^{+0.49}_{-0.46}}$&$\left(2.89^{+2.26}_{-1.46}\right)\times10^{-2}$
\\
$\eta^2_{\alpha_1\alpha_2\alpha_3}$&$3.40$-$3.60$&$29.0\pm3.0$&$40$-$47$&${4.50^{+0.45}_{-0.41}}$&$\left(3.37^{+2.47}_{-1.69}\right)\times10^{-2}$
\\
$\eta^3_{\alpha_1\alpha_2\alpha_3}$&$3.63$-$4.00$&$35.0\pm3.0$&$40$-$46$&$5.75^{+0.21}_{-0.14}$&$\left(11.71^{+4.53}_{-3.07}\right)\times10^{-2}$
\\
$\eta^4_{\alpha_1\alpha_2\alpha_3}$&$3.65$-$4.05$&$35.0\pm3.0$&$40$-$47$&$5.71^{+0.21}_{-0.14}$&$\left(16.07^{+6.09}_{-4.35}\right)\times10^{-2}$
\\
$\eta^5_{\alpha_1\alpha_2\alpha_3}$&$3.57$-$3.78$&$32.0\pm3.0$&$40$-$45$&$5.12^{+0.28}_{-0.28}$&$\left(7.15^{+4.16}_{-3.20}\right)\times10^{-2}$
\\
$\eta^6_{\alpha_1\alpha_2\alpha_3}$&$3.35$-$3.80$&$32.0\pm3.0$&$40$-$53$&$5.08^{+0.27}_{-0.28}$&$\left(10.10^{+5.68}_{-4.46}\right)\times10^{-2}$
\\ \hline
$J^1_{\alpha_1\alpha_2\alpha_3}$&$3.40$-$3.61$&$30.0\pm3.0$&$40$-$47$&${4.67^{+0.51}_{-0.42}}$&$\left(2.87^{+2.31}_{-1.49}\right)\times10^{-2}$
\\
$J^2_{\alpha_1\alpha_2\alpha_3}$&$3.39$-$3.59$&$29.0\pm3.0$&$40$-$47$&${4.49^{+0.45}_{-0.41}}$&$\left(3.35^{+2.43}_{-1.67}\right)\times10^{-2}$
\\ \hline\hline
\end{tabular}
\label{tab:results}
\end{center}
\end{table*}

In this paper we apply the QCD sum rule method to study the charmonium-like states with the exotic quantum number $J^{PC} = 3^{-+}$. Their quark contents are $q c \bar q \bar c$ ($c=charm$ and $q=up/down$), and their corresponding interpolating currents are composed of two quark fields and two antiquark fields as well as one covariant derivative operator. There are altogether six diquark-antidiquark currents, as defined in Eqs.~(\ref{def:eta1}-\ref{def:eta6}). To reach $J^{PC} = 3^{-+}$, the derivative can only be inside the diquark or antidiquark:
\begin{equation}
\eta = \big[c {\overset{\leftrightarrow}{D}} q \big] \big[\bar{c}  \bar{q} \big] + \big[c q \big] \big[\bar{c} {\overset{\leftrightarrow}{D}} \bar{q} \big] \, .
\end{equation}
We use these diquark-antidiquark currents to perform QCD sum rule analyses. The obtained results are summarized in Table~\ref{tab:results}, and the mass extracted from the current $\eta^{2}_{\alpha_1\alpha_2\alpha_3}$ is the lowest
\begin{equation}
\nonumber M_2 = {4.50^{+0.45}_{-0.41}}{\rm~GeV} \, .
\end{equation}
We have studied the mixing of $\eta^{1}_{\alpha_1\alpha_2\alpha_3}$ and $\eta^{2}_{\alpha_1\alpha_2\alpha_3}$. The obtained results are also summarized in Table~\ref{tab:results}, and the mass extracted from the mixing current $J^2_{\alpha_1\alpha_2\alpha_3}$ is slightly reduced from the single current $\eta^2_{\alpha_1\alpha_2\alpha_3}$ to be
\begin{equation}
\nonumber M_2^\prime = {4.49^{+0.45}_{-0.41}} {\rm~GeV} \, .
\end{equation}
This value is quite close to the $D^* \bar D_2^*$ threshold. Note that the authors of Refs.~\cite{Zhu:2013sca,Dong:2021juy} have applied the one-boson-exchange model to predict the existence of the $D^* \bar D_2^*$ molecular state with $J^{PC} = 3^{-+}$.

In this paper we have also constructed six meson-meson currents, as defined in Eqs.~(\ref{def:xi1}-\ref{def:xi6}). Three of them have the quark combination $[\bar c c][\bar q q]$ with the derivative between the two quark-antiquark pairs,
\begin{equation}
\xi = \big[ \bar c c \big] {\overset{\leftrightarrow}{D}} \big[ \bar q q \big] \, ;
\end{equation}
and the other three have $[\bar c q][\bar q c]$ with the derivative inside the quark-antiquark pairs,
\begin{equation}
\xi^\prime = \big[\bar c {\overset{\leftrightarrow}{D}} q \big] \big[\bar q c\big] - \big[\bar c q\big] \big[\bar q {\overset{\leftrightarrow}{D}} c\big] \, .
\end{equation}
Accordingly, a special decay behavior of the $q c \bar q \bar c$ tetraquark states with $J^{PC} = 3^{-+}$ is that: a) they decay into the $P$-wave $(\bar c c)_{S{\rm-wave}}(\bar q q)_{S{\rm-wave}}$ final states but not into the $S$-wave $(\bar c c)_{S{\rm-wave}}(\bar q q)_{P{\rm-wave}}$ and $(\bar c c)_{P{\rm-wave}}(\bar q q)_{S{\rm-wave}}$ final states, and b) they decay into the $S$-wave $(\bar c q)_{S{\rm-wave}}(\bar q c)_{P{\rm-wave}}$ final states but not into the $P$-wave $(\bar c q)_{S{\rm-wave}}(\bar q c)_{S{\rm-wave}}$ final states. Since we do not differentiate the up and down quarks in the calculations, the isospin can not be differentiated in the present study. Hence, more specifically, a) these states decay into the $P$-wave $[\rho J/\psi]/[\omega J/\psi]$ channels but not into the $S$-wave $[\rho \chi_{c2}]/[\omega \chi_{c2}]/[J/\psi f_2(1270)]$ channels, and b) they decay into the $S$-wave $[D^* \bar D_2^*]$ channel but not into the $P$-wave $[D^* \bar D^*]$ channel. Accordingly, we propose to investigate the $P$-wave $[\rho J/\psi]/[\omega J/\psi]$ channels in the future BESIII, Belle-II, and LHCb experiments to search for the charmonium-like states with the exotic quantum number $J^{PC} = 3^{-+}$.

%
\section*{Acknowledgments}
%

This project is supported by
the National Natural Science Foundation of China under Grants No.~12075019 and No.~12175318,
the Jiangsu Provincial Double-Innovation Program under Grant No.~JSSCRC2021488,
and
the Fundamental Research Funds for the Central Universities.
TGS is grateful for research funding from the Natural Sciences and Engineering Research Council of Canada (NSERC).

\appendix
\begin{widetext}
\section{Spectral densities}
\label{app:ope}
In this appendix we list the OPE spectral densities $\rho_{22\cdots66}(s)$ extracted from the currents $\eta^{2\cdots6}_{\alpha_1\alpha_2\alpha_3}$. In the following expressions, $\mathcal{F}(s) = \left[(\alpha + \beta) m_c^2 - \alpha \beta s\right]$ and $\mathcal{H}(s) = \left[m_c^2 - \alpha(1 - \alpha)s\right]$; the integration limits are $\alpha_{\min} = \frac{1 - \sqrt{1 - 4m_c^2 / s}}{2}$, $\alpha_{\max} = \frac{1 + \sqrt{1 - 4m_c^2 / s}}{2}$, $\beta_{\min} = \frac{\alpha m_c^2}{\alpha s - m_c^2}$, and $\beta_{\max} = 1 - \alpha$. The OPE spectral density $\rho_{22}(s)$ extracted from the current $\eta^2_{\alpha_1\alpha_2\alpha_3}$ is
\begin{equation}
	\rho_{22}(s) = \rho^{pert}_{22}(s) + \rho^{\qq}_{22}(s) + \rho^{\GGa}_{22}(s)+ \rho^{\qGqa}_{22}(s) + \rho^{\qq^2}_{22}(s)  + \rho^{\qq\qGqa}_{22}(s)+\rho^{\qGqa^2}_{22}(s) \, ,
	\label{ope:eta2}
\end{equation}
where
\begin{eqnarray}
    \nonumber \rho^{pert}_{22}(s) &=& \dab \Bigg\{ -\mathcal{F}(s)^5 \times \frac{1}{7680 \pi ^6 \alpha ^4 \beta ^4} \times (3 \alpha ^5+33 \alpha ^4 \beta -4 \alpha ^4+102 \alpha ^3 \beta ^2-38 \alpha ^3 \beta
    \non
    &+&138 \alpha ^2 \beta ^3-90 \alpha ^2 \beta ^2+87 \alpha  \beta ^4-82 \alpha  \beta ^3+\alpha +21 \beta ^5-26 \beta ^4+5 \beta)\Bigg\}  \, ,
    \non
    \rho^{\qq}_{22}(s) &=& {\qq } \dab \Bigg\{ m_c \mathcal{F}(s)^3  \times  \frac{-6 \alpha ^3-24 \alpha ^2 \beta +6 \alpha ^2-26 \alpha  \beta ^2+15 \alpha  \beta -8 \beta ^3+7 \beta ^2}{48 \pi ^4 \alpha ^3 \beta ^2}  \Bigg\} \, ,
    \non
    \rho^{\GGa}_{22}(s) &=& {\GGb }  \int^{\alpha_{max}}_{\alpha_{min}}d\alpha  \Bigg\{ \int^{\beta_{max}}_{\beta_{min}}d\beta \Bigg\{\mathcal{F}(s)^2 \times \frac{1}{221184 \pi ^6 \alpha ^4 \beta ^4} \times ( -\mathcal{F}(s) \times (36 \alpha ^6 \beta +288 \alpha ^5 \beta ^2-37 \alpha ^5 \beta
    \non
    &+&684 \alpha ^4 \beta ^3-342 \alpha ^4 \beta ^2-28 \alpha ^4 \beta +756 \alpha ^3 \beta ^4-918 \alpha ^3 \beta ^3+57 \alpha ^3 \beta ^2+180 \alpha ^3 \beta +432 \alpha ^2 \beta ^5-802 \alpha ^2 \beta ^4
    \non
    &+&261 \alpha ^2 \beta ^3+360 \alpha ^2 \beta ^2-236 \alpha ^2 \beta +108 \alpha  \beta ^6-189 \alpha  \beta ^5+104 \alpha  \beta ^4+180 \alpha  \beta ^3-270 \alpha  \beta ^2+85 \alpha  \beta)
    \non
    &+&24 m_c^2  (-6 \alpha ^8-54 \alpha ^7 \beta +8 \alpha ^7-150 \alpha ^6 \beta ^2+60 \alpha ^6 \beta -\alpha ^6-186 \alpha ^5 \beta ^3+126 \alpha ^5 \beta ^2-3 \alpha ^5 \beta +3 \alpha ^5
    \non
    &-&120 \alpha ^4 \beta ^4+104 \alpha ^4 \beta ^3-3 \alpha ^4 \beta ^2+6 \alpha ^4 \beta -5 \alpha ^4-78 \alpha ^3 \beta ^5+46 \alpha ^3 \beta ^4-\alpha ^3 \beta ^3+3 \alpha ^3 \beta ^2-9 \alpha ^3 \beta +\alpha ^3
    \non
    &-&90 \alpha ^2 \beta ^6+54 \alpha ^2 \beta ^5-66 \alpha  \beta ^7+60 \alpha  \beta ^6-18 \beta ^8+22 \beta ^7-4 \beta ^4)  ) \Bigg\} + \mathcal{H}(s)^3 \times \frac{5}{36864 \pi ^6 (\alpha -1) \alpha ^2} \Bigg\} \, ,
    \non
    \rho^{\qGqa}_{22}(s) &=& {\qGqb } \int^{\alpha_{max}}_{\alpha_{min}}d\alpha  \Bigg\{ \int^{\beta_{max}}_{\beta_{min}}d\beta \Bigg\{
    m_c \mathcal{F}(s)^2 \times  \frac{30 \alpha ^3-84 \alpha ^2 \beta -30 \alpha ^2-432 \alpha  \beta ^2+27 \alpha  \beta -202 \beta ^3+57 \beta ^2}{1536 \pi ^4 \alpha ^2 \beta ^2}  \Bigg\}
    \non
    &+& m_c \mathcal{H}(s)^2 \times  \frac{58 \alpha +29}{1536 \pi ^4 \alpha ^2}  \Bigg\} \, ,	
    \non
    \rho^{\qq^2}_{22}(s)&=& {\qq^2 } \int^{\alpha_{max}}_{\alpha_{min}}d\alpha \Bigg\{ m_c^2 \mathcal{H}(s) \times  \frac{1}{6 \pi ^2 \alpha}  \Bigg\} \, ,
    \non
    \rho^{\qq\qGqa}_{22}(s)&=& {\qq\qGqb } \Bigg\{ \int^{\alpha_{max}}_{\alpha_{min}}d\alpha \Bigg\{  \frac{m_c^2 (230 \alpha ^2-259 \alpha -5)}{576 \pi ^2 \alpha}  \Bigg\}  + \int^{1}_{0} \delta\left( s - {m_c^2 \over (1-\alpha) \alpha} \right) d\alpha  \Bigg\{ -\frac{m_c^4}{12 \pi ^2 \alpha}  \Bigg\}  \Bigg\} \, ,
    \non
    \rho^{\qGqa^2}_{22}(s)&=&{\qGqb^2 } \int^{1}_{0} \delta\left( s - {m_c^2 \over (1-\alpha) \alpha} \right) d\alpha  \Bigg\{ \frac{m_c^2 (316 \alpha ^3-690 \alpha ^2+359 \alpha +20)}{2304 \pi ^2 (\alpha -1) \alpha} + \frac{m_c^4 (-182 \alpha ^2+211 \alpha +5) }{ 2304 \pi ^2 (\alpha -1) \alpha ^2 M_B^2}
    \non
    &+& \frac{m_c^6} {96 \pi ^2 (\alpha -1) \alpha ^2 M_B^4} \Bigg\} \, .
\end{eqnarray}

The OPE spectral density $\rho_{33}(s)$ extracted from the current $\eta^3_{\alpha_1\alpha_2\alpha_3}$ is
\begin{equation}
	\rho_{33}(s) = \rho^{pert}_{33}(s) + \rho^{\qq}_{33}(s) + \rho^{\GGa}_{33}(s)+ \rho^{\qGqa}_{33}(s) + \rho^{\qq^2}_{33}(s)  + \rho^{\qq\qGqa}_{33}(s)+\rho^{\qGqa^2}_{33}(s) \, ,
	\label{ope:eta3}
\end{equation}
where
\begin{eqnarray}
	 \nonumber \rho^{pert}_{33}(s) &=& \dab \Bigg\{ -\mathcal{F}(s)^5 \times \frac{1}{15360 \pi ^6 \alpha ^4 \beta ^4} \times (3 \alpha ^5+33 \alpha ^4 \beta -4 \alpha ^4+102 \alpha ^3 \beta ^2-38 \alpha ^3 \beta
	 \non
	 &+&138 \alpha ^2 \beta ^3-90 \alpha ^2 \beta ^2+87 \alpha  \beta ^4-82 \alpha  \beta ^3+\alpha +21 \beta ^5-26 \beta ^4+5 \beta)\Bigg\}  \, ,
	 \non
	 \rho^{\qq}_{33}(s) &=& {\qq } \dab \Bigg\{ -m_c \mathcal{F}(s)^3  \times  \frac{-6 \alpha ^3-24 \alpha ^2 \beta +6 \alpha ^2-26 \alpha  \beta ^2+15 \alpha  \beta -8 \beta ^3+7 \beta ^2}{96 \pi ^4 \alpha ^3 \beta ^2}  \Bigg\} \, ,
	 \non
	 \rho^{\GGa}_{33}(s) &=& {\GGb }  \int^{\alpha_{max}}_{\alpha_{min}}d\alpha  \Bigg\{ \int^{\beta_{max}}_{\beta_{min}}d\beta \Bigg\{\mathcal{F}(s)^2 \times \frac{1}{221184 \pi ^6 \alpha ^4 \beta ^4} \times ( \mathcal{F}(s) \times (36 \alpha ^6 \beta +288 \alpha ^5 \beta ^2-127 \alpha ^5 \beta
	 \non
	 &+&684 \alpha ^4 \beta ^3-738 \alpha ^4 \beta ^2+82 \alpha ^4 \beta +756 \alpha ^3 \beta ^4-1170 \alpha ^3 \beta ^3+333 \alpha ^3 \beta ^2+432 \alpha ^2 \beta ^5-838 \alpha ^2 \beta ^4+321 \alpha ^2 \beta ^3
	 \non
	 &+&10 \alpha ^2 \beta +108 \alpha  \beta ^6-279 \alpha  \beta ^5+142 \alpha  \beta ^4+12 \alpha  \beta ^2-\alpha  \beta) +24 m_c^2  (-3 \alpha ^8-27 \alpha ^7 \beta +4 \alpha ^7-75 \alpha ^6 \beta ^2
	 \non
	 &+&30 \alpha ^6 \beta -93 \alpha ^5 \beta ^3+63 \alpha ^5 \beta ^2-60 \alpha ^4 \beta ^4+52 \alpha ^4 \beta ^3-\alpha ^4-39 \alpha ^3 \beta ^5+23 \alpha ^3 \beta ^4-3 \alpha ^3 \beta -45 \alpha ^2 \beta ^6+27 \alpha ^2 \beta ^5
	 \non
	 &-&33 \alpha  \beta ^7+30 \alpha  \beta ^6-9 \beta ^8+11 \beta ^7-2 \beta ^4) ) \Bigg\} + \mathcal{H}(s)^3 \times \frac{1}{36864 \pi ^6 (\alpha -1) \alpha ^2} \Bigg\} \, ,
	 \non
	 \rho^{\qGqa}_{33}(s) &=& {\qGqb } \int^{\alpha_{max}}_{\alpha_{min}}d\alpha  \Bigg\{ \int^{\beta_{max}}_{\beta_{min}}d\beta \Bigg\{
	 -m_c \mathcal{F}(s)^2 \times  \frac{6 \alpha ^3-60 \alpha ^2 \beta -6 \alpha ^2-216 \alpha  \beta ^2+27 \alpha  \beta -98 \beta ^3+33 \beta ^2}{1536 \pi ^4 \alpha ^2 \beta ^2}  \Bigg\}
	 \non
	 &-& m_c \mathcal{H}(s)^2 \times  \frac{26 \alpha +13}{1536 \pi ^4 \alpha ^2}  \Bigg\} \, ,	
	 \non
	 \rho^{\qq^2}_{33}(s)&=& {\qq^2 } \int^{\alpha_{max}}_{\alpha_{min}}d\alpha \Bigg\{ m_c^2 \mathcal{H}(s) \times  \frac{1}{12 \pi ^2 \alpha}  \Bigg\} \, ,
	 \non
	 \rho^{\qq\qGqa}_{33}(s)&=& {\qq\qGqb } \Bigg\{ \int^{\alpha_{max}}_{\alpha_{min}}d\alpha \Bigg\{  \frac{m_c^2 (118 \alpha ^2-131 \alpha -1)}{576 \pi ^2 \alpha}  \Bigg\}  + \int^{1}_{0} \delta\left( s - {m_c^2 \over (1-\alpha) \alpha} \right) d\alpha  \Bigg\{ -\frac{m_c^4}{24 \pi ^2 \alpha}  \Bigg\}  \Bigg\} \, ,
	 \non
	 \rho^{\qGqa^2}_{33}(s)&=&{\qGqb^2 } \int^{1}_{0} \delta\left( s - {m_c^2 \over (1-\alpha) \alpha} \right) d\alpha  \Bigg\{ \frac{m_c^2 (164 \alpha ^3-354 \alpha ^2+187 \alpha +4)}{2304 \pi ^2 (\alpha -1) \alpha}  + \frac{m_c^4 (-94 \alpha ^2+107 \alpha +1) }{2304 \pi ^2 (\alpha -1) \alpha ^2 M_B^2}
	 \non
	 &+& \frac{m_c^6} {192 \pi ^2 (\alpha -1) \alpha ^2 M_B^4} \Bigg\} \, .
\end{eqnarray}

The OPE spectral density $\rho_{44}(s)$ extracted from the current $\eta^4_{\alpha_1\alpha_2\alpha_3}$ is
\begin{equation}
	\rho_{44}(s) = \rho^{pert}_{44}(s) + \rho^{\qq}_{44}(s) + \rho^{\GGa}_{44}(s)+ \rho^{\qGqa}_{44}(s) + \rho^{\qq^2}_{44}(s)  + \rho^{\qq\qGqa}_{44}(s)+\rho^{\qGqa^2}_{44}(s) \, ,	
	\label{ope:eta4}
\end{equation}
where
\begin{eqnarray}
	\nonumber \rho^{pert}_{44}(s) &=& \dab \Bigg\{ -\mathcal{F}(s)^5 \times \frac{1}{7680 \pi ^6 \alpha ^4 \beta ^4} \times (3 \alpha ^5+33 \alpha ^4 \beta -4 \alpha ^4+102 \alpha ^3 \beta ^2-38 \alpha ^3 \beta
	\non
	&+&138 \alpha ^2 \beta ^3-90 \alpha ^2 \beta ^2+87 \alpha  \beta ^4-82 \alpha  \beta ^3+\alpha +21 \beta ^5-26 \beta ^4+5 \beta)\Bigg\}  \, ,
	\non
	\rho^{\qq}_{44}(s) &=& {\qq } \dab \Bigg\{ -m_c \mathcal{F}(s)^3  \times  \frac{-6 \alpha ^3-24 \alpha ^2 \beta +6 \alpha ^2-26 \alpha  \beta ^2+15 \alpha  \beta -8 \beta ^3+7 \beta ^2}{48 \pi ^4 \alpha ^3 \beta ^2}  \Bigg\} \, ,
	\non
	\rho^{\GGa}_{44}(s) &=& {\GGb }  \int^{\alpha_{max}}_{\alpha_{min}}d\alpha  \Bigg\{ \int^{\beta_{max}}_{\beta_{min}}d\beta \Bigg\{\mathcal{F}(s)^2 \times \frac{1}{221184 \pi ^6 \alpha ^4 \beta ^4} \times ( \mathcal{F}(s) \times (-36 \alpha ^6 \beta -288 \alpha ^5 \beta ^2+37 \alpha ^5 \beta
	\non
	&-&684 \alpha ^4 \beta ^3+342 \alpha ^4 \beta ^2-22 \alpha ^4 \beta -756 \alpha ^3 \beta ^4+918 \alpha ^3 \beta ^3-207 \alpha ^3 \beta ^2-432 \alpha ^2 \beta ^5+802 \alpha ^2 \beta ^4-411 \alpha ^2 \beta ^3
	\non
	&+&26 \alpha ^2 \beta -108 \alpha  \beta ^6+189 \alpha  \beta ^5-154 \alpha  \beta ^4+60 \alpha  \beta ^2-5 \alpha  \beta) +24 m_c^2  (-6 \alpha ^8-54 \alpha ^7 \beta +8 \alpha ^7-150 \alpha ^6 \beta ^2
	\non
	&+&60 \alpha ^6 \beta -186 \alpha ^5 \beta ^3+126 \alpha ^5 \beta ^2-120 \alpha ^4 \beta ^4+104 \alpha ^4 \beta ^3-2 \alpha ^4-78 \alpha ^3 \beta ^5+46 \alpha ^3 \beta ^4-6 \alpha ^3 \beta -90 \alpha ^2 \beta ^6
	\non
	&+&54 \alpha ^2 \beta ^5-66 \alpha  \beta ^7+60 \alpha  \beta ^6-18 \beta ^8+22 \beta ^7-4 \beta ^4) ) \Bigg\} 	+\mathcal{H}(s)^3 \times \frac{5}{36864 \pi ^6 (\alpha -1) \alpha ^2} \Bigg\} \, ,	
	\non
	\rho^{\qGqa}_{44}(s) &=& {\qGqb } \int^{\alpha_{max}}_{\alpha_{min}}d\alpha  \Bigg\{ \int^{\beta_{max}}_{\beta_{min}}d\beta \Bigg\{
	m_c \mathcal{F}(s)^2 \times  \frac{-30 \alpha ^3+84 \alpha ^2 \beta +30 \alpha ^2+432 \alpha  \beta ^2-27 \alpha  \beta +202 \beta ^3-57 \beta ^2}{1536 \pi ^4 \alpha ^2 \beta ^2}  \Bigg\}
	\non
	&+& m_c \mathcal{H}(s)^2 \times  \frac{-58 \alpha -29}{1536 \pi ^4 \alpha ^2}  \Bigg\} \, ,	
	\non
	\rho^{\qq^2}_{44}(s)&=& {\qq^2 } \int^{\alpha_{max}}_{\alpha_{min}}d\alpha \Bigg\{ m_c^2 \mathcal{H}(s) \times  \frac{1}{6 \pi ^2 \alpha}  \Bigg\} \, ,
	\non
	\rho^{\qq\qGqa}_{44}(s)&=& {\qq\qGqb } \Bigg\{ \int^{\alpha_{max}}_{\alpha_{min}}d\alpha \Bigg\{  \frac{m_c^2 (230 \alpha ^2-259 \alpha -5)}{576 \pi ^2 \alpha}  \Bigg\}  + \int^{1}_{0} \delta\left( s - {m_c^2 \over (1-\alpha) \alpha} \right) d\alpha  \Bigg\{ -\frac{m_c^4}{12 \pi ^2 \alpha}  \Bigg\}  \Bigg\} \, ,
	\non
	\rho^{\qGqa^2}_{44}(s)&=&{\qGqb^2 } \int^{1}_{0} \delta\left( s - {m_c^2 \over (1-\alpha) \alpha} \right) d\alpha  \Bigg\{ \frac{m_c^2 (316 \alpha ^3-690 \alpha ^2+359 \alpha +20)}{2304 \pi ^2 (\alpha -1) \alpha}  + \frac{m_c^4 (-182 \alpha ^2+211 \alpha +5) }{2304 \pi ^2 (\alpha -1) \alpha ^2 M_B^2}
	\non
	&+& \frac{m_c^6} {96 \pi ^2 (\alpha -1) \alpha ^2 M_B^4} \Bigg\} \, .
\end{eqnarray}

The OPE spectral density $\rho_{55}(s)$ extracted from the current $\eta^5_{\alpha_1\alpha_2\alpha_3}$ is
\begin{equation}
	\rho_{55}(s) = \rho^{pert}_{55}(s) + \rho^{\qq}_{55}(s) + \rho^{\GGa}_{55}(s)+ \rho^{\qGqa}_{55}(s) + \rho^{\qq^2}_{55}(s)  + \rho^{\qq\qGqa}_{55}(s)+\rho^{\qGqa^2}_{55}(s) \, ,	
	\label{ope:eta5}
\end{equation}
where
\begin{eqnarray}
	\nonumber \rho^{pert}_{55}(s) &=& \dab \Bigg\{ -\mathcal{F}(s)^5 \times \frac{1}{7680 \pi ^6 \alpha ^4 \beta ^4} \times (3 \alpha ^5+33 \alpha ^4 \beta -4 \alpha ^4+102 \alpha ^3 \beta ^2-38 \alpha ^3 \beta
	\non
	&+&138 \alpha ^2 \beta ^3-90 \alpha ^2 \beta ^2+87 \alpha  \beta ^4-82 \alpha  \beta ^3+\alpha +21 \beta ^5-26 \beta ^4+5 \beta)\Bigg\}  \, ,
	\non
	\rho^{\qq}_{55}(s) &=& 0 \, ,
	\non
	\rho^{\GGa}_{55}(s) &=& {\GGb }  \int^{\alpha_{max}}_{\alpha_{min}}d\alpha  \Bigg\{ \int^{\beta_{max}}_{\beta_{min}}d\beta \Bigg\{-\mathcal{F}(s)^2 \times \frac{1}{110592 \pi ^6 \alpha ^4 \beta ^4} \times ( \mathcal{F}(s) \times (21 \alpha ^6 \beta +195 \alpha ^5 \beta ^2+10 \alpha ^5 \beta
	\non
	&+&606 \alpha ^4 \beta ^3-26 \alpha ^4 \beta ^2-20 \alpha ^4 \beta +786 \alpha ^3 \beta ^4-282 \alpha ^3 \beta ^3-42 \alpha ^3 \beta ^2+429 \alpha ^2 \beta ^5-290 \alpha ^2 \beta ^4+30 \alpha ^2 \beta ^3-13 \alpha ^2 \beta
	\non
	&+&75 \alpha  \beta ^6-44 \alpha  \beta ^5+4 \alpha  \beta ^4-19 \alpha  \beta ^2+2 \alpha  \beta) +24 m_c^2 (3 \alpha ^8+27 \alpha ^7 \beta -4 \alpha ^7+75 \alpha ^6 \beta ^2-30 \alpha ^6 \beta +93 \alpha ^5 \beta ^3
	\non
	&-&63 \alpha ^5 \beta ^2+60 \alpha ^4 \beta ^4-52 \alpha ^4 \beta ^3+\alpha ^4+39 \alpha ^3 \beta ^5-23 \alpha ^3 \beta ^4+3 \alpha ^3 \beta +45 \alpha ^2 \beta ^6-27 \alpha ^2 \beta ^5+33 \alpha  \beta ^7-30 \alpha  \beta ^6
	\non
	&+&9 \beta ^8-11 \beta ^7+2 \beta ^4) ) \Bigg\} +\mathcal{H}(s)^3 \times \frac{1}{18432 \pi ^6 (\alpha -1) \alpha ^2} \Bigg\} \, ,	
	\non
	\rho^{\qGqa}_{55}(s) &=& 0 \, ,
	\non
	\rho^{\qq^2}_{55}(s)&=& {\qq^2 } \int^{\alpha_{max}}_{\alpha_{min}}d\alpha \Bigg\{ m_c^2 \mathcal{H}(s) \times  \frac{1}{6 \pi ^2 \alpha}  \Bigg\} \, ,
	\non
	\rho^{\qq\qGqa}_{55}(s)&=& {\qq\qGqb } \Bigg\{ \int^{\alpha_{max}}_{\alpha_{min}}d\alpha \Bigg\{  \frac{m_c^2 (28 \alpha ^2-32 \alpha -1)}{72 \pi ^2 \alpha}  \Bigg\}  + \int^{1}_{0} \delta\left( s - {m_c^2 \over (1-\alpha) \alpha} \right) d\alpha  \Bigg\{ -\frac{m_c^4}{12 \pi ^2 \alpha}  \Bigg\}  \Bigg\} \, ,
	\non
	\rho^{\qGqa^2}_{55}(s)&=&{\qGqb^2 } \int^{1}_{0} \delta\left( s - {m_c^2 \over (1-\alpha) \alpha} \right) d\alpha  \Bigg\{ \frac{m_c^2 (38 \alpha ^3-84 \alpha ^2+43 \alpha +4)}{288 \pi ^2 (\alpha -1) \alpha}  + \frac{m_c^4 (-22 \alpha ^2+26 \alpha +1) }{288 \pi ^2 (\alpha -1) \alpha ^2 M_B^2}
	\non
	&+& \frac{m_c^6} {96 \pi ^2 (\alpha -1) \alpha ^2 M_B^4} \Bigg\} \, .
\end{eqnarray}

The OPE spectral density $\rho_{66}(s)$ extracted from the current $\eta^6_{\alpha_1\alpha_2\alpha_3}$ is
\begin{equation}
	\rho_{66}(s) = \rho^{pert}_{66}(s) + \rho^{\qq}_{66}(s) + \rho^{\GGa}_{66}(s)+ \rho^{\qGqa}_{66}(s) + \rho^{\qq^2}_{66}(s)  + \rho^{\qq\qGqa}_{66}(s)+\rho^{\qGqa^2}_{66}(s) \, ,	
	\label{ope:eta6}
\end{equation}
where
\begin{eqnarray}
	\nonumber \rho^{pert}_{66}(s) &=& \dab \Bigg\{ -\mathcal{F}(s)^5 \times \frac{1}{3840 \pi ^6 \alpha ^4 \beta ^4} \times (3 \alpha ^5+33 \alpha ^4 \beta -4 \alpha ^4+102 \alpha ^3 \beta ^2-38 \alpha ^3 \beta
	\non
	&+&138 \alpha ^2 \beta ^3-90 \alpha ^2 \beta ^2+87 \alpha  \beta ^4-82 \alpha  \beta ^3+\alpha +21 \beta ^5-26 \beta ^4+5 \beta)\Bigg\}  \, ,
	\non
	\rho^{\qq}_{66}(s) &=& 0 \, ,
	\non
	\rho^{\GGa}_{66}(s) &=& {\GGb }  \int^{\alpha_{max}}_{\alpha_{min}}d\alpha  \Bigg\{ \int^{\beta_{max}}_{\beta_{min}}d\beta \Bigg\{-\mathcal{F}(s)^2 \times \frac{1}{110592 \pi ^6 \alpha ^4 \beta ^4} \times ( \mathcal{F}(s) \times (-111 \alpha ^6 \beta -753 \alpha ^5 \beta ^2+242 \alpha ^5 \beta
	\non
	&-&1074 \alpha ^4 \beta ^3+1166 \alpha ^4 \beta ^2-100 \alpha ^4 \beta -606 \alpha ^3 \beta ^4+894 \alpha ^3 \beta ^3-210 \alpha ^3 \beta ^2-447 \alpha ^2 \beta ^5+326 \alpha ^2 \beta ^4+150 \alpha ^2 \beta ^3
	\non
	&-&41 \alpha ^2 \beta -273 \alpha  \beta ^6+356 \alpha  \beta ^5+20 \alpha  \beta ^4-95 \alpha  \beta ^2+10 \alpha  \beta) +48 m_c^2 (3 \alpha ^8+27 \alpha ^7 \beta -4 \alpha ^7+75 \alpha ^6 \beta ^2-30 \alpha ^6 \beta
	\non
	&+&93 \alpha ^5 \beta ^3-63 \alpha ^5 \beta ^2+60 \alpha ^4 \beta ^4-52 \alpha ^4 \beta ^3+\alpha ^4+39 \alpha ^3 \beta ^5-23 \alpha ^3 \beta ^4+3 \alpha ^3 \beta +45 \alpha ^2 \beta ^6-27 \alpha ^2 \beta ^5+33 \alpha  \beta ^7
	\non
	&-&30 \alpha  \beta ^6+9 \beta ^8-11 \beta ^7+2 \beta ^4) ) \Bigg\} +\mathcal{H}(s)^3 \times \frac{5}{18432 \pi ^6 (\alpha -1) \alpha ^2} \Bigg\} \, ,
	\non
	\rho^{\qGqa}_{66}(s) &=& 0 \, ,
	\non
	\rho^{\qq^2}_{66}(s)&=& {\qq^2 } \int^{\alpha_{max}}_{\alpha_{min}}d\alpha \Bigg\{ m_c^2 \mathcal{H}(s) \times  \frac{1}{3 \pi ^2 \alpha}  \Bigg\} \, ,
	\non
	\rho^{\qq\qGqa}_{66}(s)&=& {\qq\qGqb } \Bigg\{ \int^{\alpha_{max}}_{\alpha_{min}}d\alpha \Bigg\{  \frac{m_c^2 (62 \alpha ^2-67 \alpha +1)}{72 \pi ^2 \alpha}  \Bigg\}  + \int^{1}_{0} \delta\left( s - {m_c^2 \over (1-\alpha) \alpha} \right) d\alpha  \Bigg\{ -\frac{m_c^4}{6 \pi ^2 \alpha}  \Bigg\}  \Bigg\} \, ,
	\non
	\rho^{\qGqa^2}_{66}(s)&=&{\qGqb^2 } \int^{1}_{0} \delta\left( s - {m_c^2 \over (1-\alpha) \alpha} \right) d\alpha  \Bigg\{ \frac{m_c^2 (88 \alpha ^3-186 \alpha ^2+101 \alpha -4)}{288 \pi ^2 (\alpha -1) \alpha}  + \frac{m_c^4 (-50 \alpha ^2+55 \alpha -1) }{288 \pi ^2 (\alpha -1) \alpha ^2 M_B^2}
	\non
	&+& \frac{m_c^6} {48 \pi ^2 (\alpha -1) \alpha ^2 M_B^4} \Bigg\} \, .
\end{eqnarray}	
\end{widetext}

\bibliographystyle{elsarticle-num}
\bibliography{ref}

\end{document}